\documentclass[%
reprint,
groupedaddress,
 amsmath,amssymb,
 aip,
]{revtex4-1}

\usepackage{graphicx}
\begin{document}
%\begin{center} 
\title{Suspending superconducting qubits by silicon micromachining}

\author{Y. Chu}
\email{yiwen.chu@yale.edu}
\author{C. Axline}
\author{C. Wang}
\author{T. Brecht}
\author{Y. Y. Gao}
\author{L. Frunzio}
% \author{M.H. Devoret}
\author{R.J. Schoelkopf}
\affiliation{Department of Applied Physics, Yale University, New Haven, Connecticut 06511, USA}

\date{\today}

\begin{abstract}

% We present a method for improving the lifetimes of aluminum 3D transmon qubits on silicon using micromachining. Our technique is a one-step deep reactive ion etch that requires no additional fabrication processes and has high yield. The surface loss of the qubits is reduced by suspending the junction area and the edges of the aluminum film. We are able to obtain $T_1$'s of more than 60 $\mu$s, comparable to state-of-the-art qubits made on sapphire. Similarly to the case of aluminum on sapphire, we find that the lifetimes of conventional silicon qubits are dominated by surface loss. Our suspended structures, however, appear to be limited by other mechanisms such as bulk dielectric loss. The drastic change in the device geometry enabled by our technique can also be used to study other effects of the environment on qubit behavior. As an example, we find that suspension increases the flux noise experienced by tunable SQUID-based qubits.  

We present a method for relieving aluminum 3D transmon qubits from a silicon substrate using micromachining. Our technique is a high yield, one-step deep reactive ion etch that requires no additional fabrication processes, and results in the suspension of the junction area and edges of the aluminum film. The drastic change in the device geometry affects both the dielectric and flux noise environment experienced by the qubit. In particular, the participation ratios of various dielectric interfaces are significantly modified, and suspended qubits exhibited longer $T_1$'s than non-suspended ones. We also find that suspension increases the flux noise experienced by tunable SQUID-based qubits.

% \begin{description}
% Secondary publications and information retrieval purposes.
% \item[PACS numbers]
% May be entered using the \verb+\pacs{#1}+ command.
% You may use the \texttt{description} environment to structure your abstract;
% use the optional argument of the \verb+\item+ command to give the category of each item. 
% \end{description}

\end{abstract}

% \keywords{microwave resonator, cavity resonator, superconducting, micromachining, bulk etching, indium}%Use showkeys class option if keyword
\maketitle

%\tableofcontents

% \section{Introduction}

The coherence times of superconducting qubits have steadily increased over the past decade due to careful engineering of the electromagnetic environment, better materials and fabrication methods, and improved device designs that minimize loss. State-of-the-art superconducting qubits with the longest lifetimes ($T_1$) make use of very low loss tangent dielectric substrates and have large separation between planar conductors to decrease the effect of dielectric loss in the interfaces between materials \cite{Paik:2011hd,BarendsPRL2013,MartinisArxiv2014}. In particular, it has been shown that for aluminum 3D transmons on sapphire, $T_1$ times are limited by the various interfaces between the dielectric substrate, the superconducting metal, and vacuum \cite{Wang2015}. This effect can be attributed to the larger electric fields near metallic surfaces and the higher concentration of two-level systems (TLS) at disordered interfaces \cite{Martinis2005, ChangAPL2013, GaoAPL2008}. At the same time, magnetic impurities at the surface of superconductors have been proposed as the cause of $1/f$ flux noise that limits the performance of SQUID based qubits and sensors \cite{YoshiharaPRL2006, Bialczak2007, Anton2013}.

In order to better understand these effects, one strategy is to drastically alter the geometry of materials and interfaces that contribute to qubit loss and decoherence. In this Letter, we present a procedure for removing the substrate and suspending aluminum Josephson junctions on silicon by micromachining. Silicon is a low-loss dielectric that offers several advantages for implementing the next generation of complex quantum circuits \cite{OConnellAPL2008, GambettaArxiv2016}. Its prevalent use in the semiconductor and MEMS industries have led to a large variety of fabrication techniques that are not available for sapphire \cite{KovacsIEEE1998}. Using silicon as a substrate material enables the development of novel devices and architectures in circuit QED, such as multilayer quantum circuits that incorporate micromachined superconducting enclosures and resonators \cite{Brecht:NPJ, Brecht:APL}. Substrate micromachining has also been used to reduce dielectric loss and frequency noise in niobium titanium nitride coplanar waveguide resonators on silicon \cite{Bruno:2015up, BarendsAPL2010}. On the other hand, silicon has a more complex surface chemistry than sapphire; for example, it forms an amorphous oxide layer that may be host to a large number of TLS's and paramagnetic impurities \cite{OConnellAPL2008, VonSchickfusPLA1977,Koch2007}. 

We suspend our qubits with a simple, one-step deep reactive ion etch (DRIE) using the BOSCH process that does not require any additional steps to mask or protect the devices \cite{LaermerPatent1996, Bruno:2015up, SUPPLEMENT}. We begin with high resistivity (100) silicon wafers ($\rho>10^4$ $\Omega\cdot$cm) and fabricate aluminum 3D transmon qubits using the standard Dolan bridge double-angle deposition technique\cite{DolanAPL1997}. DRIE is then performed directly on the fabricated qubits. This is possible because aluminum itself is an excellent mask for the BOSCH process. We have performed this process on more than one hundred devices of various geometries, and found that the Josephson junctions were almost all unaffected except for a slight increase in the normal state resistance, possibly due to increased diffusion of the junction oxide when the devices are heated by the etching process. 

\begin{figure}
%\onecolumn
%\linespread{1}
%\begin{center}
\includegraphics[scale = 1,angle=0]{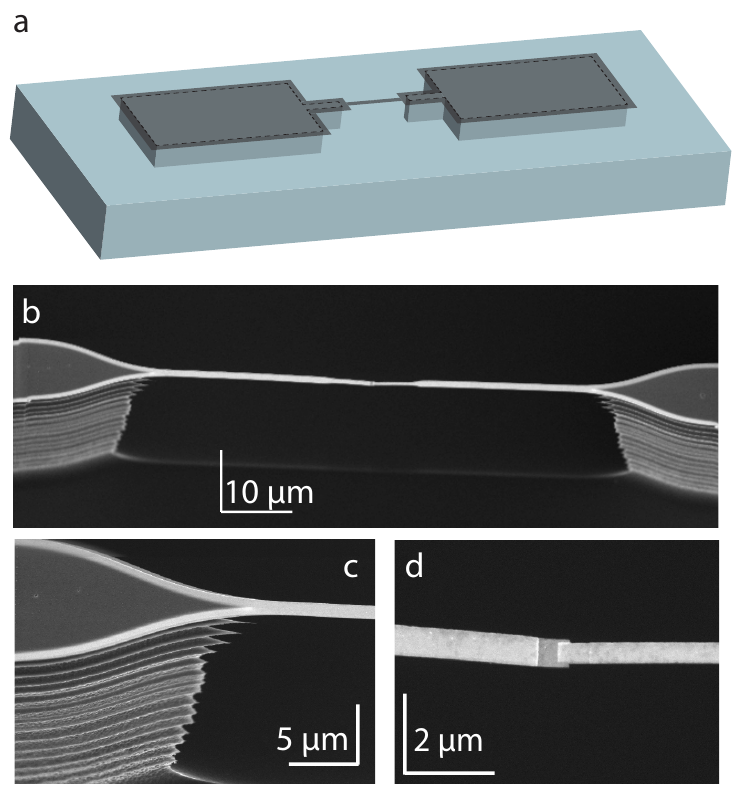}
\caption{
	\textbf{Micromachined 3D transmons using DRIE}  
	\textbf{(a)} A schematic drawing of a suspended 3D transmon on silicon. An overhang is created at the edges of the metal, while thin features are suspended. 
    \textbf{(b)} SEM image of a BOSCH etched transmon showing the entire suspended leads around the junction region
    \textbf{(c)} Detailed view of the supporting silicon pedestals, showing the lighter-colored overhanging Al edges and corrugated BOSCH profile.
    \textbf{(e)} Detailed view of the completely suspended Al-AlO$_x$-Al Josephson junction.
}	
\label{Fig1}
%\end{center}
\end{figure}

Figure \ref{Fig1} shows a schematic drawing and scanning electron micrographs (SEM) of suspended 3D transmons. We note that all regions $\sim$700 nm from the edge are undercut, resulting in the junction region and the narrow 1 $\mu$m wide leads on either side becoming completely suspended. This means that our process is compatible with aluminum based devices of any geometry, as long as suspended metal regions are supported by other larger features. The suspended single junction transmons are robust against solvent cleaning, drying, and repeated thermal cycles. We observe, however, that more complex suspended structures such as SQUID loops are more easily damaged, for example by surface tension during solvent cleaning or wet etches.

\begin{figure}[ht]
%\twocolumn
%\linespread{1}
%\begin{center}
\includegraphics[scale = 1.1, angle=0]{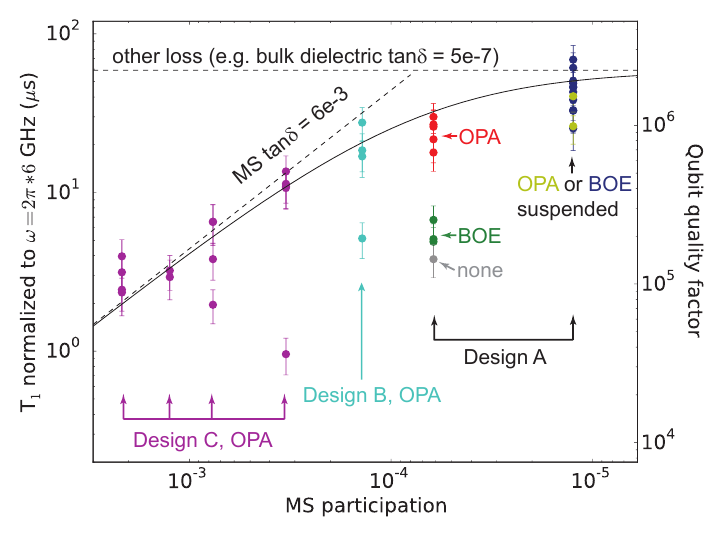}
\caption{
	\textbf{Lifetime of silicon transmons vs. metal-substrate participation ratio} Each point corresponds to one measured qubit. The MS participation of design C qubits were varied by changing planar capacitor gap distance \cite{SUPPLEMENT}. The label OPA signifies oxygen plasma ashed before and after deposition, while BOE signifies buffered oxide etch before deposition only. Errorbars are typical variation of $T_1$'s over time. Dashed lines are guides to the eye corresponding MS surface loss and other effects that are independent of MS participation, such as bulk dielectric loss. Solid line indicates the combinination of these loss mechanisms.
%     Dashed line corresponds to the bulk silicon loss tangent limit suggested by the data, using a constant bulk participation of 90$\%$. 
}	
\label{Fig2}
%\end{center}
\end{figure}

We first study the effect our process has on dielectric loss and qubit $T_1$. Following the analysis in \textcite{Wang2015}, we quantify the loss due to various dielectric materials using their participation ratios and loss tangents (tan$\delta$). In Figure \ref{Fig2}, we plot the measured $T_1$'s of several types of qubits with different designs and fabrication procedures against the simulated participation ratios of their metal-substrate (MS) interfaces. Similar plots of $T_1$ versus the substrate-air (SA) and metal-air (MA) participation ratios can be found in the supplementary materials \cite{SUPPLEMENT}. In addition to qubits of the design shown in Figure \ref{Fig1} (Design A), we also fabricated a set of qubit designs with higher surface participation (Designs B and C). Design C qubits have planar capacitors whose gaps can be varied to change the surface participations. Drawings of all qubit designs can be found in the supplementary materials \cite{SUPPLEMENT}. The dielectric participation ratios were obtained through electromagnetic simulations that faithfully modeled the device geometries, including the undercut at the edge of suspended devices \cite{Wang2015, SUPPLEMENT}. 

Without any kind of surface preparation before or after aluminum deposition, the typical $T_1$'s of the Design A qubits are only a few microseconds, which is more than an order of magnitude worse than the same design on sapphire. The use of surface treatment techniques such as buffered oxide etch (BOE) or oxygen plasma ashing (OPA) \cite{SUPPLEMENT, SlichterThesis2011} improves the lifetimes of the regular non-suspended qubits.
% , with the latter producing a MS loss tangent of tan$\delta \sim 7\times 10^{-3}$ as indicated by the various unetched designs. 
DRIE further improves the $T_1$'s of the Design A qubits. 
% The best results were obtained when BOE or OPA was performed before aluminum deposition, followed by OPA before and after BOSCH etching.
The highest $T_1$ measured with this procedure was $\sim$63 $\mu$s for a etch depth of 60 $\mu$m, which is comparable to the $T_1$'s of typical qubits of the same design on sapphire. We note that the qubit loss is likely frequency dependent because of coupling to resonant loss channels such as TLS's. This can lead to low $T_1$'s in exceptional cases, which we have also included in Figure \ref{Fig2}. We will explore this further in our later discussion of flux tunable qubits.

The results in Figure \ref{Fig2} indicate that the quality of interfaces of qubits on silicon are highly dependent on surface treatments and are generally higher loss than those on sapphire. The $T_1$'s of non-suspended Design A qubits suggest that surface treatment before and after deposition is important on silicon. On the other hand, only OPA before deposition is needed to obtain $>$50 $\mu$s $T_1$'s on sapphire \cite{Wang2015}. It is possible that, for example, the liftoff process leaves more resist residue on silicon than on sapphire. In addition, any exposed silicon surface will form an oxide even after cleaning. Both resist and oxide are likely to result in a higher loss SA interface, which has comparable participation ratios as the MS interface for non-suspended qubits. This may explain the observation that while qubit $T_1$'s are better after surface cleaning, they never reach the levels measured on sapphire. 
% We point out that the the $T_1$'s of suspended qubits are independent of whether or not OPA was performed before etching \cite{SUPPLEMENT}. This is consistent with the fact that the original SA surface is completely etched away. 

It is also evident from Figure \ref{Fig2} that it is insufficient to consider a simple model where qubit loss is dominated by single dielectric interface. The qubits with high MS participation (Design C) have $T_1$'s that follow a line of constant MS tan$\delta = 6\times10^{-3}$, consistent with being limited by loss due to that interface. However, qubits with lower MS participation (Designs A and B) deviate from that line. One simple explanation for this trend is that the bulk dielectric loss for our silicon substrates becomes significant once the MS participation has been sufficiently reduced. The bulk dielectric participations of the measured qubits, including the suspended ones, are all similar to within 10$\%$. Therefore, we can indicate the $T_1$ limit due to bulk dielectric loss as a horizontal line in Figure \ref{Fig2}. We find that taking into account both bulk and MS surface loss mechanisms results in a model that is consistent with data from both the suspended qubits and the regular qubits that underwent OPA. 

We emphasize, however, that other loss mechanisms can play a role as well. For example, the SA and MA participations scale similarly to the MS participation for the regular, non-suspended qubits\cite{Wang2015}. However, unlike a change in qubit geometry, the DRIE process affects the three interfaces differently. In particular, it increases both the SA and MA participations \cite{SUPPLEMENT}. Therefore, increased loss from the SA and MA interfaces could contribute to negating the $T_1$ improvement expected from the reduced MS participation. We also cannot rule out more complex effects of the DRIE, such as a change in the SA and MA tan$\delta$'s from damage or polymer deposition on these surfaces. Clearly, more investigations are needed to isolate and understand these different effects. Our results indicate that micromachining is a new technique to alter the loss contributions of various materials in ways not possible with changes in geometry alone. This can help us gain information about the roles of individual interfaces in limiting qubit $T_1$'s.

\begin{figure*}[ht]
%\twocolumn
%\linespread{1}
%\begin{center}
\includegraphics[scale = 1.2,angle=0]{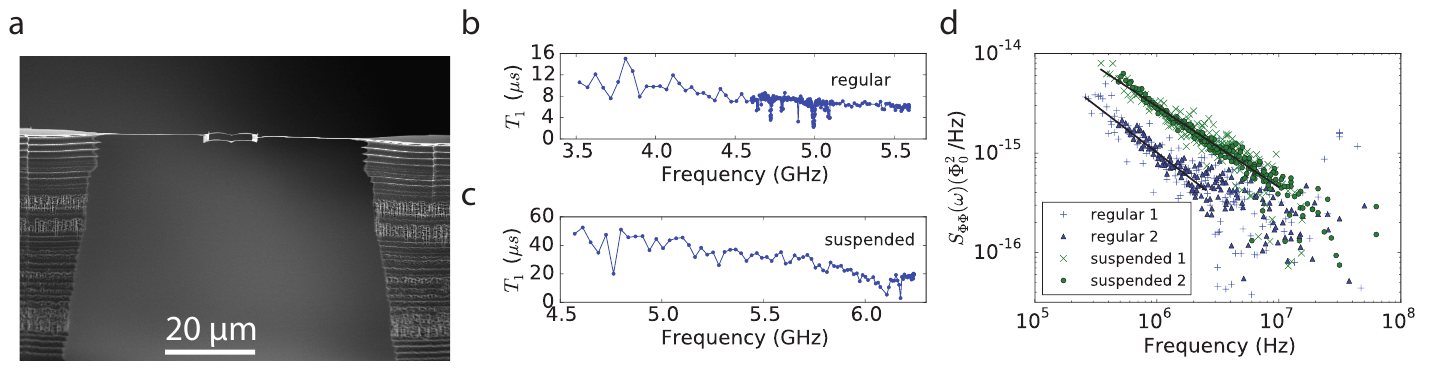}
\caption{
	\textbf{Flux tunable suspended qubits}  
	\textbf{(a)} Side view SEM image of a suspended SQUID transmon. \textbf{(b, c)} $T_1$ vs qubit frequency for regular \textbf{(b)} and suspended \textbf{(c)} qubits. See supplementary materials \cite{SUPPLEMENT} for zoomed-in view of a low-$T_1$ feature in \textbf{(b)}. \textbf{(d)} PSD of flux noise extracted from dynamical decoupling of regular and suspended qubits, each including data from two qubits. Solid lines are fits to the data. The extent of the lines indicate the frequency range in which the data was above the noise floor and therefore included in the fit.
}	
\label{Fig3}
%\end{center}
\end{figure*}

In order to investigate the effects of DRIE on qubit behavior in more detail, we also measured frequency dependence of the $T_1$ and flux noise of regular and suspended tunable SQUID qubits. The qubit design is exactly the same as in Figure \ref{Fig1}, except the single junction is replaced by a 10 $\mu$m $\times$ 10 $\mu$m SQUID loop, which is completely suspended after etching. A side-view SEM of a suspended SQUID transmon is shown in Figure \ref{Fig3}a. We compare this device with another that underwent BOE before deposition and no other surface cleaning or etching after deposition. The two qubits were symmetrically arranged inside the same copper cavity to ensure that they experienced a similar background electromagnetic environment. Two separate solenoid coils mounted outside the cavity and aligned with the location of the qubits allowed us to individually control the frequency of each device. 

We plot $T_1$ as a function of qubit frequency for one pair of regular and suspended qubits in Figure \ref{Fig3}b and \ref{Fig3}c, respectively. For the regular qubit, we find that, in addition to a low overall $T_1$ of $<$10 $\mu$s, there are sharp dips in the $T_1$ at a multitude of distinct frequencies. The suspended qubit shows a higher overall $T_1$, but also exhibits a few resonant features where the the $T_1$ is drastically reduced. A second pair of qubits measured in the same manner exhibited similar behavior. These observations imply that both types of qubits are affected by the presence of resonant loss channels such as TLS's, as was observed in many previous studies \cite{Martinis2005, Lisenfeld2015}. We observe that resonant loss channels are less prevalent for the suspended qubit than the regular qubit. However, their presence may explain the variability that we observe in the $T_1$ measurements of single junction qubits. 

The 3D SQUID transmons also allow us to investigate if and how suspension affects their magnetic environments. In particular, many previous works have observed $1/f$ flux noise experienced by several different types of superconducting qubits \cite{YoshiharaPRL2006, Bialczak2007, Anton2013}. It has been proposed that, similarly to dielectric loss, flux noise can be caused by defects in amorphous surface materials, such as silicon oxide \cite{Koch2007}. However, typical measured flux noise levels in SQUIDs have been orders of magnitude higher than estimates based on known sources, and the origin of this important dephasing mechanism for superconducting qubits remains uncertain \cite{Bylander2011, Bialczak2007, Anton2013}. 

In order to measure the noise power spectral density (PSD) of the qubits, we use the technique demonstrated in \textcite{Bylander2011}. We measure the response of the qubits to a collection of Carr-Purcell-Meiboom-Gill (CPMG) dynamical decoupling sequences with varying time delays and number of pulses to filter out noise at different frequencies. We can then use this in combination with the measured frequency-flux curves for the qubits to extract the PSD within a range of noise frequencies \cite{Bylander2011, SUPPLEMENT}. The results are shown in Figure \ref{Fig3}d. The PSD's include data from two pairs of regular and suspended qubits measured in the same cavity on successive cooldowns. The data from the two pairs of qubits agree very well with each other, indicating that observed differences between the two types of qubits are not due to sample to sample variations. The flux noise for both qubits exhibits a clear power law dependence with an exponent of $\alpha = 0.8\pm 0.2$ for the suspended qubit and $\alpha = 0.9\pm 0.4$ for the regular qubit. We find that, while the PSD of both qubits is consistent with the range of previous measurements, the flux noise of the suspended qubits is higher than the regular qubits by a factor of of $\sim$3 in the frequency range measured. We also performed the same measurements at the zero-flux points of each qubit, and found that the PSD was essentially flat at the level of $10^{-16}$ $\Phi_0^2/$Hz. This suggests that the measured noise in this frequency range is likely to be flux-related and not due to, for example, critical current fluctuations.

While one might expect that removing the substrate from underneath the SQUID loop would decrease the flux noise due to surface spins, our data indicates that the opposite effect occurs. It has been suggested that the dominant contributors to flux inside the SQUID loop are spins on the surface of the loop traces \cite{Koch2007}. Therefore, removing the silicon surface inside and outside the loop may not have a large effect on the flux noise. On the other hand, DRIE also exposes the bottom surface of the aluminum loop, which forms a layer of amorphous AlO$_x$ in air. Our observations are consistent with the new AlO$_x$ layer having a higher concentration of spins than the aluminum-silicon interface, possibly because most of the SiO$_x$ was removed by BOE prior to deposition. The observation that the flux noise increased by more than a factor of two after etching could suggest that the new AlO$_x$ layer on the bottom surface contains more defects than the top surface. This might be the case given that the top oxide layer was grown in pure oxygen conditions inside the evaporator rather than through exposure to air \cite{Schneider1999}. We emphasize that while this explanation is consistent with our observations, further investigation would be needed to elucidate the microscopic origins of additional flux noise in suspended qubits.

We have demonstrated that micromachining of silicon substrates is compatible with aluminum Josephson junction qubits. The process results in a reduction of the metal-substrate interface and an improvement of qubit $T_1$'s. Our results seem to suggest that we are approaching a regime where qubit decay is dominated by other mechanisms such as dielectric loss of the bulk silicon substrate. The loss tangent of ``undoped" high-resistivity silicon is not very well known or understood, and is likely to be dependent on residual dopants and defects. We speculate that the DRIE technique described here, in combination with higher quality substrates, can result in qubits with even longer lifetimes. In addition, MS and bulk participation ratios can be further reduced by redesigning the qubit so that the DRIE process suspends larger areas of the device. We emphasize, however, that dielectric loss will eventually become dominated by another material. Even in the limit of a qubit floating in vacuum, there will be dielectric loss due to, for example, oxide on the surface of the metal. Beyond the reduction of dielectric loss, our measurements of flux noise with suspended SQUID transmons is another example of how qubit properties can be altered by changing the geometry of the substrate and the materials present in the environment. We expect that other potential loss mechanisms for cQED devices, such as quasiparticles and phonon coupling\cite{CatelaniPRB2011, IoffePRL2004}, will also be affected. Therefore, our process provides a tool for understanding and improving the various aspects involved in the performance of superconducting qubits.

We thank Luke Burkhart, Mollie Schwartz, and Michel Devoret for valuable discussions. Facilities use was supported by the Yale SEAS cleanroom, YINQE and NSF MRSEC DMR-1119826. This research was supported by the Army Research Office under Grant No. W911NF-14-1-0011. C.A. acknowledges support from the NSF Graduate Research Fellowship under Grant No. DGE-1122492. Y.Y.G. acknowledges support from an A*STAR NSS Fellowship.

% \bibliography{SiQubitMM}
%\bibliographystyle{unsrtnat}
%\bibliographystyle{apalike2}
%\bibliographystyle{naturemag_noURL}

% Supplementary material must be listed in the reference section as follows: ?See supplemental material at [URL will be inserted by AIP] for [give brief description of material].? 

%%%%%%%%%%%
\clearpage

%%%%%%%%%%%%%%%%%%%%%%%%%%%
%%   SUPPLEMENTARY MATERIAL
%%%%%%%%%%%%%%%%%%%%%%%%%%%

\pagebreak
\onecolumngrid
\begin{center}
\textbf{\large Supplemental Materials for "Suspending superconducting qubits by silicon micromachining"}
\end{center}

%%%%%%%%%% Prefix a "S" to all equations, figures, tables and reset the counter %%%%%%%%%%
\setcounter{equation}{0}
\setcounter{figure}{0}
\setcounter{table}{0}
\setcounter{page}{1}
\setcounter{section}{0}
\makeatletter
\renewcommand{\theequation}{S\arabic{equation}}
\renewcommand{\thefigure}{S\arabic{figure}}
\renewcommand{\thetable}{S\arabic{table}}
% \renewcommand{\bibnumfmt}[1]{[S#1]}
% \renewcommand{\citenumfont}[1]{S#1}
%%%%%%%%%%%%

\section{Qubit designs and fabrication procedures}
\begin{figure*}[ht]
%\twocolumn
%\linespread{1}
%\begin{center}
\includegraphics[scale = 2.5,angle=0]{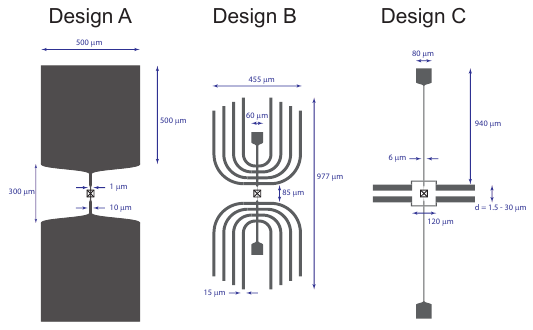}
\caption{
	\textbf{Transmon qubit geometries used in this study} The surface participations of Design C were varied by changing the distance $d$ between the planar capacitor features. 	
}	
\label{FigS1}
%\end{center}
\end{figure*}

Three types of 3D transmon qubit geometries were used to vary the surface participation in Figure 2 of the main text. They were similar to the designs used in \textcite{Wang2015}, and are shown in Figure \ref{FigS1}. Design A, which had the lowest surface participations, was used to study the effects of DRIE etching. 

Table \ref{table:fabVsT1} shows the various combinations of fabrication steps performed on Design A qubits and the resulting best $T_1$'s. Fabrication of all qubits was performed using the Dolan bridge technique on 300-500 um thick (100) high resistivity ($\rho>10^4$ $\Omega$cm) silicon wafers (Crystec and SiliconQuest). After cleaning in acetone and methanol, the wafer was spun with a bilayer of e-beam resist consisting of 500 nm of MMA (8.5) MAA EL 13 and 70 nm of 950K PMMA A3, then baked at 175 $^{\circ}$C. Patterning of the qubit was done on a 100 kV VISTEC EBPG 5000+ e-beam writer, and the wafer was subsequently developed for 55 seconds in 1:3 MIBK:IPA followed by a 5 second rinse in IPA. For some of the devices, the developed wafer was then either etched in 10:1 BOE for 10 s or in an oxygen plasma asher (Glow Research AutoGlow) at 100 W for 10 s. We call this post-development step ``clean 1" in Table \ref{table:fabVsT1}. The wafer was then loaded into a Plassys e-beam evaporation system (MEB550S or UMS 300). In the case of BOE, the wafer was kept in a vacuum box during transport to the evaporator, which takes about five minutes. A bi-layer of aluminum (20 nm and 60 nm) was deposited using double-angle evaporation. In between the two layers, the junction barrier was grown by thermal oxidation using using a 85:15 Ar:O$_2$ mixture at 15 Torr for 12 minutes. Finally, the aluminum was capped with another oxide layer grown at 3 Torr for 10 minutes. After deposition, liftoff was performed in 90 $^{\circ}$C NMP for several hours, then rinsed with acetone and methanol. Prior to dicing into individual chips with one qubit each, a layer of photoresist was spun on the wafer to protect the qubits. After dicing in an ADT 7100 dicer, the resist was removed by rinsing in solvent. 

The diced chips were then optionally oxygen plasma ashed (OPA) at 100 W for 3 minutes (clean 2) and/or DRIE etched. The DRIE was done using the BOSCH process with alternating SF$_6$ inductively coupled plasma (ICP) etch (10 s, 35 mTorr, 700 W) and C$_4$F$_8$ ICP passivation (3s, 35 mTorr, 700 W) steps. The plasma was turned off for 1 minute after every 5 etch/passiviation cycles to prevent overheating of the sample. By changing the ratio of the etch and passivation step times, we can control the amount of undercut and the orientation of the sidewalls. A larger ratio results in a larger undercut and a sidewall that slopes more inward with etch depth. Finally, the total number cycles controls the overall depth of the etch. We find that a 10s/3s etch/passivation cycle, resulting in a $\sim$700 nm undercut, and a 60 um etch depth gives a significant modification of the participation ratios without jeopardizing the structural integrity of the devices. 

After DRIE, some of the chips were processed with OPA again at 100 W for 3 minutes to remove any deposited polymer that might remain from the BOSCH process (clean 3).

\begin{table}[h]
\centering
\caption{\textbf{Effect of cleaning procedures and etching on qubit T1s.} Parameters and placement of the clean and DRIE steps in the fabrication procedure are described in the text. BOE: Buffered oxide etch. OPA: Oxygen plasma ashing.}
\label{table:fabVsT1}
\begin{tabular}{|l|l|l|l|l|}
\hline

Clean 1		& Clean 2 	& DRIE	& Clean 3	& Max $T_1$ ($\mu$s)\\ \hline \hline
None	 	& None		& None  & None    	& 4 \\ \hline
BOE		 	& None		& None  & None  	& 6 \\ \hline
BOE		 	& None		& Yes	& None  	& 63 \\ \hline
BOE		 	& None		& Yes	& OPA	 	& 59 \\ \hline
BOE		 	& OPA		& None	& None	 	& 7 \\ \hline
BOE		 	& OPA		& Yes	& OPA	 	& 50 \\ \hline
OPA		 	& OPA		& None	& OPA	 	& 23 \\ \hline
OPA		 	& OPA		& Yes	& OPA	 	& 44 \\ \hline
       
\end{tabular}
\end{table}

\section{Simulations of dielectric participation}
\begin{figure*}[ht]
%\twocolumn
%\linespread{1}
%\begin{center}
\includegraphics[scale = 0.6,angle=0]{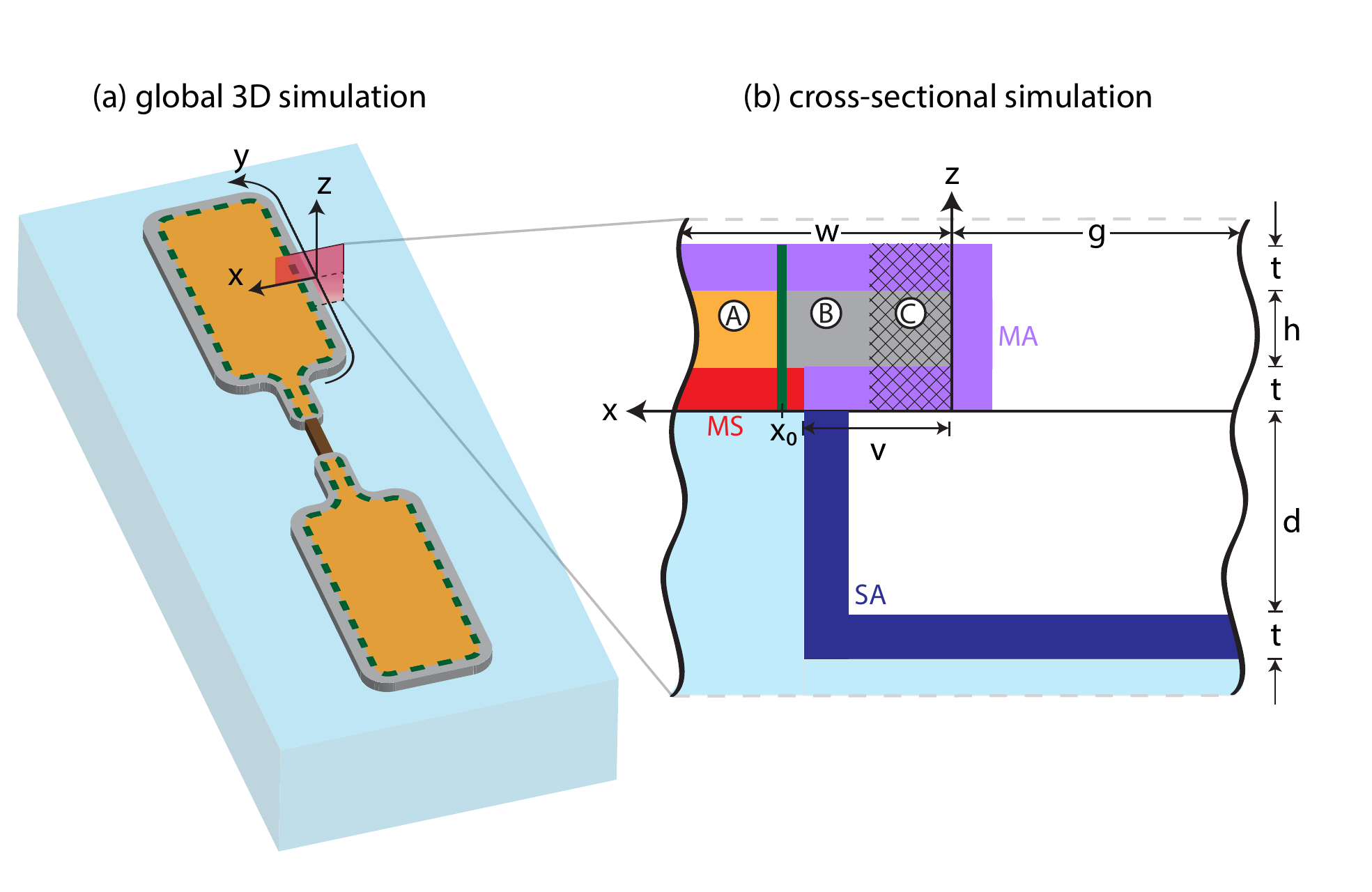}
\caption{
	\textbf{Simulation geometry. (a)} An isometric overview of the transmon pads on a silicon substrate (light blue). The leads near the junction, completely suspended in DRIE processed geometries, are shown in brown. The separation of the inner region (orange) and exterior region (gray) of the pads is indicated by a green dashed line. A slice taken on the xz plane is shaded in red. \textbf{(b)}  The slice taken in \textbf{(a)} shows the etch profile, including the metal overhang (width $v$) and deep substrate etch ($\sim d$ deep). Interface surfaces are assumed to share a common thickness $t$, taken here to be 3\,nm. The MS layer (red), which would normally extend to the edge of the metal, is reduced by the etch. The MA layer (purple) also includes the underside of any metal freshly exposed by the etch. The SA layer (dark blue) includes the sidewall surface of the remaining support substrate, increasing coverage significantly compared to unetched devices. The scalloped profile of the SA surface was approximated using a smooth plane. In accordance with the procedure in \textcite{Wang2015}, the perimeter region (gray) is divided into a cross-hatched region \textcircled{C}, which can fail to converge in the global simulation if $v$ is small compared to $x_0$. The region \textcircled{B} is convergent in both simulations, and is used to bridge them in order to calculate the energy in the layers within region \textcircled{C}. The division between \textcircled{A} and \textcircled{B}, at $x=x_0$, separates the inner and exterior regions. All dimensions are not to scale.}
\label{SimGeom}
%\end{center}
\end{figure*}

Simulations for surface and bulk dielectric participation ratios were performed a method similar to that in \textcite{Wang2015}. For the suspended qubits, the geometries of the various interfaces were modified accordingly, as shown in Figure~\ref{SimGeom}. 
%The metal-surface (MS) interface was defined as the dielectric surface in contact with the metal that remains after etching. Since electric field is concentrated towards the overhanging end of the conductor, values for MS participation are reduced in the presence of etching. The definition of the metal-air (MA) interface, which previously included dielectric surface on the top and sides of the metal, was expanded to include the area on the underside of overhanging or suspended metal. The sidewalls of the etched silicon were considered part of the surface-air (SA) dielectric layer, which included the top interface of the substrate. 
%The scalloped profile of the SA surface was approximated using a smooth plane.

\begin{figure*}[ht]
%\twocolumn
%\linespread{1}
%\begin{center}
\includegraphics[scale = 1.2,angle=0]{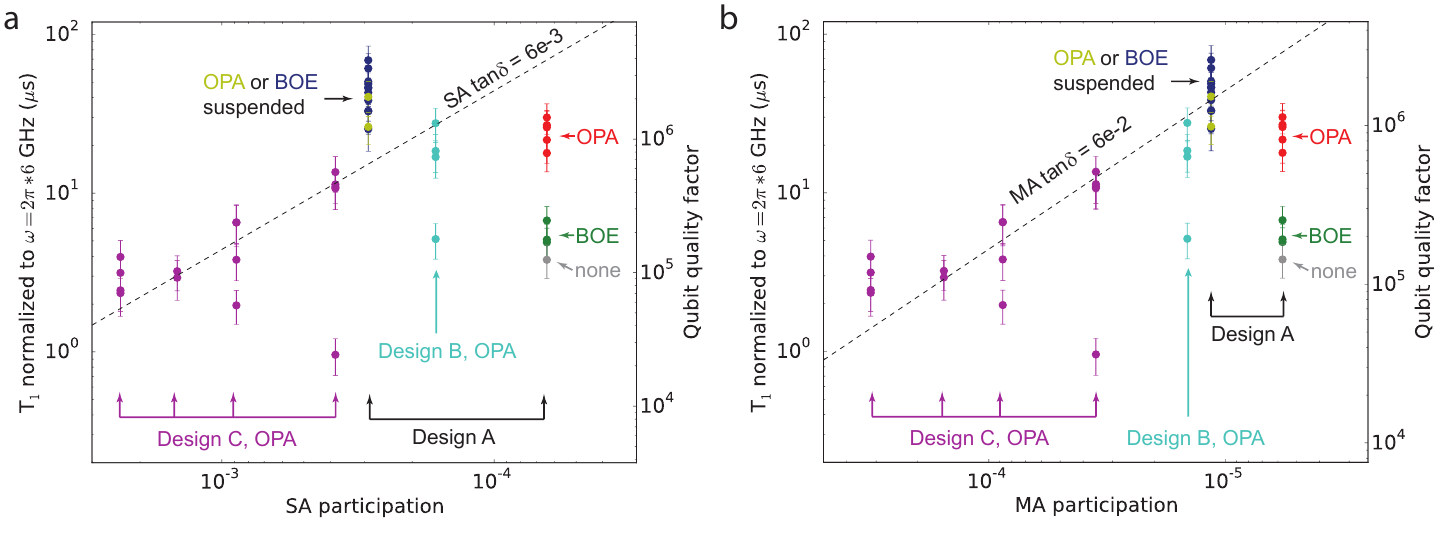}
\caption{
	\textbf{Lifetime of silicon transmons vs. SA (a) and MA (b) participation ratios.} Dashed lines are guides to the eye corresponding to the indicated dielectric loss tangents. 
}	
\label{FigS_PRs}
%\end{center}
\end{figure*}

The DRIE suspension processes affects the substrate-air (SA) and metal-air (MA) interfaces in a qualitatively different way than the metal-substrate (MS) interface. While it reduces the MS participation by about a factor of 5, it \textit{increases} the SA participation by a factor of 4.5 and the MA participation by a factor of 3, as shown in Table \ref{table:participations} and \ref{FigS_PRs}. As mentioned in the main text, the increase of these participation ratios could contribute to additional loss in the suspended qubits, which would explain why their $T_1$'s fall below the constant MS loss tangent line in Figure 2 of the main text. However, it is evident from Figure \ref{FigS_PRs} that the unetched Design A qubits have shorter $T_1$'s than would be expected from a constant MS, SA, or MA loss tangent. Therefore, as proposed in the main text, an additional loss mechanism is needed to explain this discrepancy. The simplest model assumes an effect that is independent of surface participations, such as bulk dielectric loss.

\begin{table}[h]
\centering
\caption{\textbf{Surface participation ratios for various qubits} The dimensions indicated for Design C are the gap sizes $d$ for the planar capacitors, as shown in Figure \ref{FigS1}.}
\label{table:participations}
\begin{tabular}{|l|l|l|l|l|}
\hline

Qubit	& MS 	& SA	& MA	\\ \hline \hline
Design A, suspended	 	& 1.25e-5	& 2.90e-4  & 5.74e-6   \\ \hline
Design A, regular	& 6.16e-5 &6.40e-5& 1.90e-6 \\ \hline
Design B, & 1.39e-4&1.64e-4&1.45e-5\\ \hline
Design C, 30 $\mu$m	& 3.32e-4&3.83e-4&3.53e-5 \\ \hline
Design C, 20 $\mu$m	& 3.96e-4&4.55e-4&4.22e-5\\ \hline
Design C, 10 $\mu$m	& 5.63e-4&6.49e-4&6.19e-5 \\ \hline
Design C, 6 $\mu$m	& 7.64e-4&8.85e-4&8.74e-5 \\ \hline
Design C, 3 $\mu$m	& 1.25e-3&1.46e-3&1.55e-4\\ \hline
Design C, 1.5 $\mu$m	& 2.16e-3&2.36e-3&3.16e-4 \\ \hline
       
\end{tabular}
\end{table}

\section{Resonant loss features}
\begin{figure*}[ht]
%\twocolumn
%\linespread{1}
%\begin{center}
\includegraphics[scale = 1.2,angle=0]{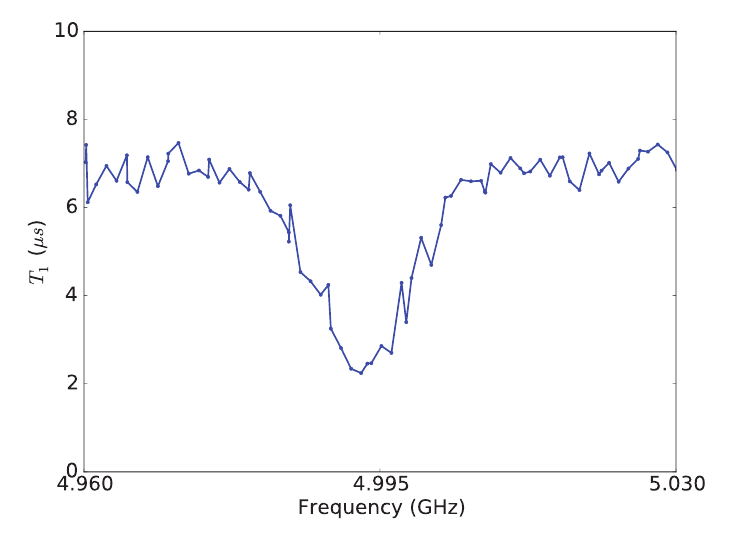}
\caption{
	\textbf{Resonant loss feature in non-suspended SQUID qubit} Zoom-in on one of the resonant features in Figure 3b of the main text. 
}	
\label{FigS_T1Dip}
%\end{center}
\end{figure*}

Both suspended and non-suspended flux tunable SQUID qubits exhibited dips in $T_1$ at certain frequencies, consistent with resonant loss mechanisms usually attributed to TLS's in previous works \cite{Martinis2005,Lisenfeld2015}. They appear to be more prevalent in the non-suspended qubits, which exhibited several of such features of various widths. A detailed measurement of $T_1$ versus frequency around one of these features is shown in \ref{FigS_T1Dip}.

\section{Flux noise analysis}

The power spectral densities (PSD) of the flux noise experienced by suspended and regular SQUID transmons were measured using dynamical decoupling techniques as described in \textcite{Bylander2011}. We briefly summarize the procedure here. 

We applied a series of Carr-Purcell-Meiboom-Gill (CPMG) sequences \cite{MeiboomRSI1958}, given by 
\begin{equation}
\left(\frac{\pi}{2}\right)_x - \left[\frac{\tau}{2N} - \pi_y - \frac{\tau}{2N}  \right]^N - \left(\frac{\pi}{2}\right)_x 
\end{equation}

followed by dispersive measurement of the qubit state. The qubits were tuned to some flux point near $\Phi_0/4$ where the $T_1$ was relatively long. In our case, the measured signal is given by

\begin{equation}
\chi_N(\tau) = A_0+A \left(\frac{\partial \omega_q}{\partial \Phi}\right)^2 \tau^2 \int_0^\infty d\omega S_{\Phi\Phi}(\omega)g_N (\omega, \tau)
\label{EqCPMGSig}
\end{equation}

Where $\omega_q$ is the qubit frequency, $A_0$ and $A$ are an overall offset and scaling determined by the qubit readout parameters, and $S_{\Phi\Phi}(\omega)$ is the PSD of the flux noise. By varying the length of the total sequence $\tau$ and the number of echo pulses $N$, we can vary the center frequency and bandwidth of the filter function $g_N (\omega, \tau)$ that makes the dynamical decoupling sequence sensitive to a particular part of the noise spectrum \cite{AlvarezPRL2011, UhrigPRL2007}. The expression for $g_N (\omega, \tau)$ for the CPMG sequence can be found in the supplementary materials for \textcite{Bylander2011}. 

\begin{figure*}[ht]
%\twocolumn
%\linespread{1}
%\begin{center}
\includegraphics[scale = 1.2,angle=0]{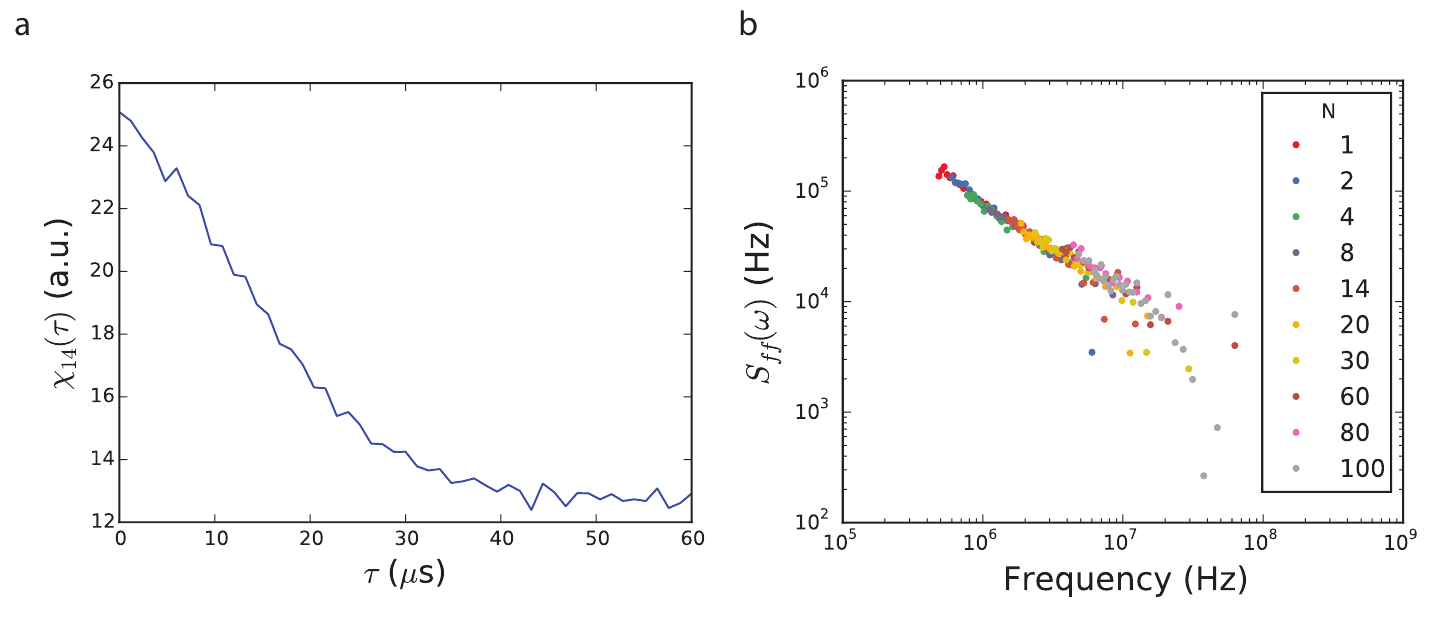}
\caption{
	\textbf{Determination of flux noise PSD} \textbf{(a)} Raw CPMG data with $N = 14$. \textbf{(b)} PSD of frequency noise for a suspended qubit including data for all different $N$'s.
}	
\label{FigS3}
%\end{center}
\end{figure*}

Our goal is to invert equation \ref{EqCPMGSig} in order to determine $S_{\Phi\Phi}(\omega)$ from the raw data. Each set of raw data consists of varying $\tau$ for a fix number of echo pulses, an example of which is shown in Figure \ref{FigS3}a. We then fit the data to a functional form
\begin{equation}
A_0+A e^{-\frac{(\tau- \tau_0)^2}{2T_2^2}} e^{-\frac{\tau-\tau_0}{T_1}},
\label{EqFitFxn}
\end{equation}
which takes into account the independently measured $T_1$ decay of the qubit and assumes a particular functional form for the dephasing, which of course depends on the flux noise PSD that we would like to determine. Therefore, the fit is only used to extract $A_0$, $A$, and $T_2$, which are used to normalize the raw data and determine the range of $\tau$ for which the data is above the noise floor. 

Next, we simplify the problem by assuming that $g_N (\omega, \tau)$ is sharply peaked and replace it with a rectangular filter function with the same height and width. This becomes a better approximation for larger $N$ and $\tau$. Using this simplified filter function, we can calculate the power spectral density of the \textit{frequency} noise. An example is shown in Figure \ref{FigS3}b after combining datasets with different $N$'s for one of the suspended qubits. Note that each data point in the raw data corresponds to one data point in the PSD. Finally, to convert this into flux noise, we determine $\partial \omega_q/\partial \Phi$ by measuring the flux tuning spectrum of the qubits. 

\bibliography{SiQBosch}

\end{document}